\documentstyle[12pt,epsfig]{article}

\begin{document}   

\def\beq{\begin{equation}}
\def\eeq{\end{equation}}
\def\eq{\beq\eeq}
\def\beqn{\begin{eqnarray}}
\def\eeqn{\end{eqnarray}}
\relax
\def\ba{\begin{array}}
\def\ea{\end{array}}
\def\squ{\tilde{Q}}
\def\tb{\mbox{tg$\beta$}}
\def\hH{\hat{H}}
\def\tu{\tilde{u}}
\def\td{\tilde{d}}
\def\te{\tilde{e}}
\def\tQ{\tilde{Q}}
\def\tL{\tilde{L}}
\def\tt{\tilde{t}}
\def\tb{\tilde{b}}
\def\ttau{\tilde{\tau}}
\def\tnu{\tilde{\nu}}
\def\ra{\rightarrow} 
\newcommand{\nn}{\nonumber}
\newcommand{\lsim}{\raisebox{-0.13cm}{~\shortstack{$<$ \\[-0.07cm] $\sim$}}~}
\newcommand{\gsim}{\raisebox{-0.13cm}{~\shortstack{$>$ \\[-0.07cm] $\sim$}}~}
\newcommand{\s}{\\ \vspace*{-2mm} }
\renewcommand{\theequation}{\thesection.\arabic{equation}}

\begin{flushright}
PM/98--26 \\
hep-ph/9810214
\end{flushright}

\vspace{1cm}

\begin{center}

{\large\sc {\bf 
EXTRACTING CHARGINO/NEUTRALINO MASS PARAMETERS FROM PHYSICAL OBSERVABLES\\}} 
\vspace{1cm}
{G. MOULTAKA}
\vspace{1cm}

Physique Math\'ematique et Th\'eorique, UMR--CNRS, \\
Universit\'e Montpellier II, F--34095 Montpellier Cedex 5, France.
\\E-mail: moultaka@lpm.univ-montp2.fr

\end{center}

\vspace*{2cm}

\begin{abstract}

\noindent 
I report on two papers, hep-ph/9806279 and hep-ph/9807336,
where complementary strategies are proposed for the
determination of the chargino/neutralino sector parameters, $M_1, M_2, \mu
$ and $\tan \beta$, from the knowledge of some physical observables.
This determination and the occurrence of possible ambiguities 
are studied as far as possible analytically within the context of the
unconstrained MSSM, assuming however no CP-violation.

\vspace{4cm}
\begin{center}
{\it Talk given at the International Conference
on High Energy Physics, \\Vancouver 1998\\
(to appear in the proceedings)}
\end{center}
\end{abstract}

\newpage
\section{Introduction}

The gauge bosons and Higgs bosons superpartners have every chance to play,
in the minimal version of the supersymmetric standard model (MSSM),
an important part in the first direct experimental evidence for supersymmetry,
if the latter happens to be linearly realized in nature around the
electroweak scale. This would go through the study of the direct
production of the light states and their subsequent decays, eventually
cascading down to leptons (or jets) and missing energy \cite{detection}${}^{,}$\cite{eemach1}${}^{,}$\cite{eemach2}.

The chargino/neutralino sector is an over-constrained system in the sense
that only a few basic parameters in the Lagrangian are needed to determine
all the six physical masses and the mixing angles of the various states.
The latter determine the couplings to gauge bosons, Higgs bosons and
matter fermions, so that various phenomenological tests could
be in principle envisaged in the process of experimental identification. 
Alternatively, one might hope that a partial experimental knowledge
of this sector would be sufficient to allow a reasonably unequivocal 
reconstruction of the full set of parameters; at stakes, on one hand the 
determination of the magnitude of the fermion soft susy breaking parameters,
on the other, the existence of a heavy neutral stable particle, of prime
importance to the cold dark matter issue \cite{darkmatter}.
Furthermore, the sensitivity to $\tan \beta$, the ratio of the two vacuum 
expectation values of the Higgs fields, and to the supersymmetric parameter
$\mu$, brings in a further correlation with the other sectors of the MSSM.\\

Hereafter we describe two strategies: the first deals with the extraction of
$M_2, \mu$ and $ \tan \beta$ form the study of the lightest chargino
pair production and decay in $e^+ e^-$ collisions \cite{CDDKZ}, the second
with the extraction of $M_1, M_2$ and $\mu$ form the knowledge of any three
ino masses and $\tan \beta$ \cite{KM}. 
We start by stating the common features to these
complementary approaches as well as their specific assumptions. 
We then highlight 
the main ingredients of each of them and illustrate some of their results. 
Finally we show in what sense they eventually complement one another. 
[Obviously, the reader is referred to \cite{CDDKZ} and \cite{KM} for more 
details and references. Still, we add some comments at various places of the 
ongoing presentation, which
differ slightly from, and hopefully complete, the latter references.]\\

The reconstruction of the basic parameters of the theory involves generically
two steps which can be sketched as follows:\\

\begin{equation}
\begin{array}{lcl}
&\mbox{Experimental Observables}& \\
&\Big{\updownarrow}\;\; (I) &\\
&\mbox{Physical Parameters}& \\ 
&\Big{\updownarrow}\;\; (II) & \\
&\mbox{Lagrangian parameters} &
\end{array}\label{phil}
\end{equation}
Each of these steps can suffer from equivocal reconstructions due to partial
experimental knowledge or to theoretical ambiguities. In the present report
we concentrate on the theoretical aspects of both steps.  

\section{CDDKZ and KM common features}
The ino sector is considered in both \cite{CDDKZ} 
(referred to as CDDKZ) and \cite{KM} (KM) with the following assumptions:

\begin{itemize}
\item{} No reference to 
model-dependent assumptions about physics 
at energies much higher than the electroweak scale, like the GUT scale,
and their possible implication on the parameters of this sector. 
[Thus the study is mainly carried out in the unconstrained MSSM, but
any model-assumptions can be easily overlaid.]

\item{}  R-parity conservation; 

\item{} CP-conservation in the ino sector;
This assumption is here only for practical reasons and should be eventually
removed in future studies in order to cope with the possibility 
to deal with (complex) phases \cite{phases};

\item{} CDDKZ and KM
choose $M_2>0$. This is of course a mere convention due to
the partial phase freedom through redefinition of fields,
the only physical signs being the relative ones among $M_1, M_2$ and  $\mu$ 
as one can easily see from the relevant terms in the Lagrangian. 
(also $\tan \beta$ is taken positive and the $\mu$ term convention is
that of ref.[\cite{haber+erratum}].)
\end{itemize}
Let us now recall briefly the basic ingredients of the ino mass matrices.
The physical charginos (resp. neutralinos) are mixtures of charged
(resp. neutral) higgsino and gaugino components.  
The chargino mass matrix reads:

 \begin{equation}
\label{Mchargino}
{\cal M}_C = \left(
  \begin{array}{cc} M_2  & \sqrt 2 m_W \sin\beta  \\
              \sqrt 2 m_W \cos\beta  & \mu  
\end{array} \right)
\end{equation}
It has a supersymmetric contribution coming from the $\mu$ term
in the superpotential, the higgsino component, a contribution
from the soft susy breaking wino mass term, and off-diagonal
terms due to the electroweak symmetry breaking. Since ${\cal M_C}$
is not symmetric one needs two independent unitary matrices for the 
diagonalization.
This is but the reflection of the fact that there are two independent
mixings involving separately the two higgsino $SU(2)_L$ doublets.
The eigenvalues are most easily obtained from the diagonalization
of ${\cal M_C}^\dagger {\cal M_C}$ giving the squares of the chargino
masses:   

\begin{equation}
\begin{array}{rcl}
M^2_{\chi^\pm_{1,2}}
  &=&\frac{1}{2}[M^2_2+\mu^2+2m^2_W\\
&&    \mp\sqrt{(M^2_2+\mu^2+2m^2_W)^2-4(M_2\mu-m^2_W\sin 2\beta)^2}]
\end{array}\label{chmass}
\end{equation}

On the other hand, the angles $\phi_L, \phi_R$ defining the two independent
left- and right- chiral mixings among the winos and higgsinos in the four
component Dirac representation are given by 

\begin{equation}
\begin{array}{rcl}
\cos 2 \phi_L &=& \frac{M_2^2 - \mu^2 - 2 m_W^2 \cos 2 \beta}{2( M_{\chi_1}^2 -
m_W^2) - M_2^2 -\mu^2}\\
&& \\
\sin 2 \phi_L &=& \frac{2\sqrt{2} m_W (M_2 \cos \beta + \mu \sin \beta)}{2( M_{\chi_1}^2 - m_W^2) - M_2^2 -\mu^2} \\
&& \\
\cos 2 \phi_R &=& \frac{M_2^2 - \mu^2 + 2 m_W^2 \cos 2 \beta}{2( M_{\chi_1}^2 -
m_W^2) - M_2^2 -\mu^2}\\
&& \\
\sin 2 \phi_R &=& \frac{2\sqrt{2} m_W (M_2 \sin \beta + \mu \cos \beta)}{2( M_{\chi_1}^2 - m_W^2) - M_2^2 -\mu^2} \\
\end{array}\label{mixing}
\end{equation}

where $ M_{\chi_1}$ is the lightest chargino mass given by eq.(\ref{chmass}).
This form of the mixing angles is such that the eigenvalues of ${\cal M_C}$
are always positive definite.\\ 

The neutralino mass matrix corresponds to bilinear terms in the photino, 
zino and neutral higgsino two-component fields. It receives contributions
from the $\mu$ term, the soft mass terms of the gaugino $SU(2)_L$ triplet
$(M_2)$ and singlet $(M_1)$, while the mixing among states is triggered
by the electroweak symmetry breaking:

%\begin{table}[h]
\begin{equation}
\begin{array}{lll}
{\cal M_N} &=&
\!\left(\!\!\!
  \begin{array}{cccc} M_1 & 0 & -m_Z s_w\! \cos\beta & m_Z s_w\! \sin\beta  \\
  0 & M_2 &  m_Z c_w \!\cos\beta & -m_Z c_w\! \sin\beta  \\
 -m_Z s_w \!\cos\beta & m_Z c_w \!\cos\beta & 0 & -\mu \\
m_Z s_w \!\sin\beta & -m_Z c_w \!\sin\beta & -\mu & 0 
\end{array} \!\right)  \\
\!\!\!\!\label{neutmass}
\end{array}
\end{equation}
%\end{table}

In contrast with ${\cal M_C}$, ${\cal M_N}$ is symmetric so that it can
be diagonalized with one unitary matrix. On the other hand the eigenmasses
are not positive definite\footnote{For more details about the ino sector see 
for instance \cite{haber+erratum},\cite{haberkane} and references therein.}.
Finally we note that in general the diagonalization of ${\cal M_N}$ 
cannot be achieved through a similarity transformation, 
unless all three parameters $M_1, M_2$ and $\mu$ are
real. This will be a key point in the algorithm we present for the 
reconstruction of the parameters in the neutralino sector.

\section{Specific features}
\subsection{CDDKZ} 
The lightest chargino $\chi_1^+$ can be produced in pairs in $e^+ e^-$
collisions, at LEPII \cite{eemach1} or NLC \cite{eemach2} energies, 
through $\gamma$ and $Z$ s-channel 
exchange as well as sneutrino
t-channel exchange. The production cross section will thus depend on
the chargino mass $m_{\chi_1}$, the sneutrino mass $m_{\tilde{\nu}}$ and the
mixing angles, eq.(\ref{mixing}), which determine the couplings of the chargino
states to the $Z$ and the sneutrino. The unpolarized total cross section is 
illustrated in fig.1 with three representative cases of higgsino, gaugino or
mixed content of the lightest chargino mass. The sharp rise near threshold 
should allow a precise determination of the chargino mass. 
Also the sensitivity to the sneutrino mass with the typical destructive 
interference in the gaugino and mixed cases necessitates the knowledge of 
this parameter.\\ 
Subsequently the chargino will
decay directly to a pair of matter fermions (leptons or quarks) and the 
(stable) lightest neutralino, through the exchange of a $W$ boson (charged Higgs
exchange is suppressed for light fermions) or scalar partners of leptons or 
quarks. Of course the decay matrix elements will depend on further parameters
like the susy scalar masses and couplings to the neutralino. However, CDDKZ
propose that, looking at the total production cross section and some 
polarization components and spin-spin correlations of the final state 
charginos, one can define measurable combinations for which the details of the 
chargino decay products cancel out. This allows to isolate to a large
extent the chargino system from the neutralino system and thus extract
the mixing angles and chargino mass from these observables (step (I) in 
eq.(\ref{phil})). In fact, for step (I) to work completely for the chargino
system, one needs to know, besides the sneutrino mass,
{\sl also the lightest neutralino mass}, as 
will become clear later on. Once the chargino mass $m_{\chi^+_1}$ and 
$\cos 2 \phi_L, \cos 2 \phi_R$ are known one can determine $M_2, \mu$ and 
$\tan \beta$ up to possible ambiguities, [step (II) of eq.(\ref{phil}).]\\

Before going further let us first describe in some detail the basic ingredients
of step (I). The presence of invisible neutralinos, in the 
final state of the process
$ e^+ e^- \to \chi_1^+ \chi_1^- \to \chi_1^0 \chi_1^0 
(f_1 \bar{f}_2) (\bar{f}_3 f_4)$,
makes it impossible to measure directly the chargino production 
angle in the laboratory frame. From now on we will thus concentrate on 
observables where this angle is integrated out. Integrating also over the
invariant masses of the fermionic systems $(f_1 \bar{f}_2)$ and
$(\bar{f}_3 f_4)$ one can write the differential cross section in the
following form:
   
\begin{figure}
\hspace{-1cm}
\epsfig{figure=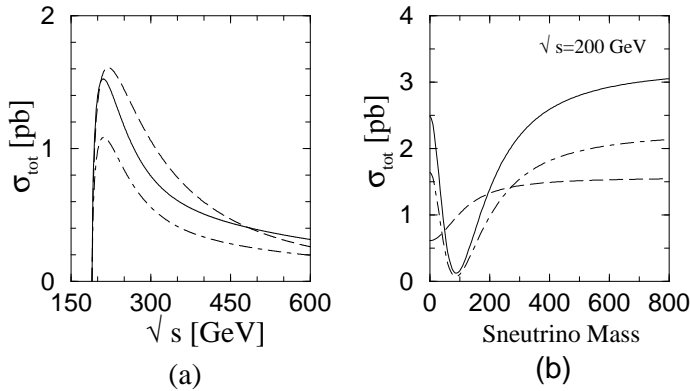, height= 2in}
\caption{Total cross section for the charginos pair production for a 
representative set of $M_2, \mu$, solid line gaugino case, dashed line
higgsino case, dot-dashed line mixed case. In (a) $m_{\tilde{\nu}}=200GeV$.
(taken from ref.[1])}
\label{fig:cddkz2}
\end{figure}

\begin{figure}
\center
\epsfig{figure=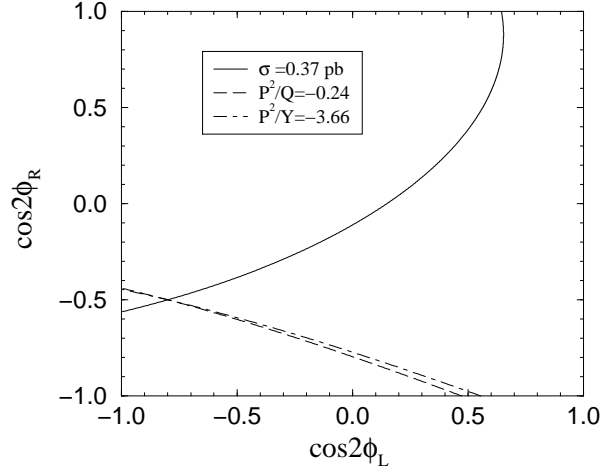, height= 3in}
\caption{Contours for the ``measured values'' of the total cross section
(solid line), $\frac{{\cal P}^2}{\cal{Q}}$, and $\frac{{\cal P}^2}{\cal{Y}}$ 
(dot-dashed line) for $m_{\chi_1^{\pm}}=95 GeV$ [$m_{\tilde{\nu}} = 250 GeV$].
(taken from ref.[1])}
\label{fig:cddkz1}
\end{figure}

%\begin{table*}[t]
\begin{equation}
\begin{array}{llr}
\frac{d^4 \sigma( e^+ e^- \to \chi_1^+ \chi_1^- \to \chi_1^0 \chi_1^0 
(f_1 \bar{f}_2) (\bar{f}_3 f_4)) }{d\cos \theta^* d\phi^* d\cos \bar{\theta}^*
d\bar{\phi}^*} &=&
\frac{\alpha^2 \beta}{124 \pi s} Br_{\chi^- \to \chi^0 f_1 \bar{f}_2}
Br_{\chi^+ \to \chi^0 \bar{f}_3 f_4} \Sigma(\theta^*, \phi^*, \bar{\theta}^*,
\bar{\phi}^*) 
\end{array}
\end{equation}
%\end{table*}
where $\alpha$ is the fine structure constant, $\beta$ the velocity of
the chargino in the c.m. frame,
$\theta^*$ ($\bar{\theta}^*$) denotes the polar angle of the 
$f_1 \bar{f}_2$ ($\bar{f}_3 f_4$) system
in the $\chi^-_1$ ($\chi^+_1$) rest frame with respect to the charginos flight 
direction in the laboratory frame, and $\phi^*$ ($\bar{\phi}^*$)
 the corresponding azimuthal 
angle with respect to a canonical production plane.  
$\Sigma(\theta^*, \phi^*, \bar{\theta}^*,
\bar{\phi}^*)$ is made out of combinations of helicity amplitudes which lead to
an unpolarized term plus fifteen other contributions from polarization
components and spin-spin correlations. We reproduce here only those components
which are relevant to our discussion.  

%\newpage
\begin{equation}
\begin{array}{rcl}
\Sigma & = & \Sigma_{unpol} + (\kappa - \bar{\kappa}) \cos \theta^* {\cal P}
+ \cos \theta^* \cos \bar{\theta^*} \kappa \bar{\kappa} {\cal Q} \\
&& \\
&& + \sin \theta^* \sin \bar{\theta^*} \cos (\phi^* + \bar{\phi^*}) \kappa 
\bar{\kappa} {\cal Y} + \dots 
\end{array} \label{Sigma}
\end{equation}

Actually, among the sixteen terms which contribute to $\Sigma$ only ten survive
because of CP-invariance (when violation of CP from the Z-boson width
or radiative corrections is neglected). Of these ten, three 
are redundant being CP eigenstates. Of the remaining seven independent 
components, only those which can be extracted from experimentally measurable 
angular distributions are explicitly written in eq.(\ref{Sigma}). 
This means that the others will be integrated out through appropriate 
projections.
 
In eq.(\ref{Sigma}) $\Sigma_{\rm unpol}$ corresponds to the unpolarized cross 
section for the chargino pair production and is given in terms of helicity
amplitudes by

\begin{equation}
\begin{array}{rcr}
\Sigma_{\rm unpol}&=&\frac{1}{4}\int {\rm d}\cos\Theta\sum_{\sigma=\pm}
      [|\langle\sigma;++\rangle|^2+|\langle\sigma;+-\rangle|^2 \\
&&           +|\langle\sigma;-+\rangle|^2+|\langle\sigma;--\rangle|^2
      ]
\end{array}
\end{equation}

${\cal P}$ is a polarization component coming separately from the $\chi^-$ 
(or $\chi^+$) system
\begin{equation}
\begin{array}{rcr}
{\cal P}&=&\frac{1}{4}\int{\rm d}\cos\Theta\sum_{\sigma=\pm}
      [|\langle\sigma;++\rangle|^2+|\langle\sigma;+-\rangle|^2 \\
       &&    -|\langle\sigma;-+\rangle|^2-|\langle\sigma;--\rangle|^2
      ]\\
\end{array}
\end{equation}

while ${\cal Q}$ and ${\cal Y}$ describe the spin correlations between the two
chargino systems and have the following structure
\begin{equation}
\begin{array}{rcr}
{\cal Q}&=&\frac{1}{4}\int {\rm d}\cos\Theta\sum_{\sigma=\pm}
      [|\langle\sigma;++\rangle|^2-|\langle\sigma;+-\rangle|^2 \\
      &&     -|\langle\sigma;-+\rangle|^2+|\langle\sigma;--\rangle|^2
      ]\\
& & \\
{\cal Y}&=&\frac{1}{2}\int {\rm d}\cos\Theta\sum_{\sigma=\pm} 
      {\rm Re}\{\langle\sigma;--\rangle\langle\sigma;++\rangle^*\}
\end{array}
\end{equation}

where $\sigma$ is the initial state electron helicity. 
The strategy of CDDKZ is based on the following two observations:
\begin{itemize}
\item[{\it i)}]
The three angular contributions, $\cos \theta^*, \cos \bar{\theta^*}$ and
$\sin \theta^* \sin \bar{\theta^*} \cos( \phi^* + \bar{\phi^*})$ are fully 
determined by the measurable parameters $E, |\vec{P}|$ (the energy and momentum
of each of the decay systems $f_i \bar{f_j}$ in the laboratory frame),
 and the chargino mass $m_{\chi_1^+}$;

\item[{\it ii)}] The three quantities $\Sigma_{unpol}, {\cal P}^2/{\cal Q}$
and ${\cal P}^2/{\cal Y}$ lead to $\kappa$ free observables,
where $\kappa$ (and $\bar{\kappa} = - \kappa)$) measures the asymmetry
between left- and right-chirality form factors in the decay products
of the chargino;
\end{itemize}

Here the kinematic configuration is similar to that of a $\tau$ lepton 
pair production with successive decays in light leptons or quarks
plus missing energy. However in the present context the invisible particle
has a non negligible mass whose knowledge is necessary to relate the energy
of the $f_i \bar{f_j}$ system in the chargino rest frame to that in the
laboratory frame. Thus the neutralino mass is actually necessary
in the reconstruction of the angular contributions in {\it i)}.
The crucial point in {\it ii)} is that the dependence on the final state
decay fermions through the asymmetry in the left- and right- chiral
structure cancels out. Thus $\Sigma_{unpol}, {\cal P}^2/{\cal Q}$
and ${\cal P}^2/{\cal Y}$ allow to study the chargino sector independently
of the details of the decay products. In the same time, their extraction
from the experimental data, via convolution with appropriate moments,
requires the measurement of the energies and momenta of the two $f_i \bar{f}_j$
systems, the chargino mass (ex. via threshold effects, see fig.1),
as well as the neutralino mass (ex. from the energy distribution of the final 
particles). Once extracted, one can combine $\Sigma_{unpol}, {\cal P}^2/{\cal Q}$ and ${\cal P}^2/{\cal Y}$ which depend on the c.m. energy $\sqrt{s}$,
the sneutrino mass $m_{\tilde{\nu}}$, and $\cos 2 \phi_L, \cos 2 \phi_R$
to determine the latter cosines. An illustration is given in fig. 2 of
a unique consistent solution corresponding to the intersection point of the
contour plots at $(\cos 2 \phi_L=-0.8, \cos 2 \phi_R=-0.5)$. The
requirement that the three curves should meet in one point offers clearly a very
stringent consistency check of the model. On the other hand, while 
$\Sigma_{unpol}$ is a quadratic polynomial in $\cos 2 \phi_L$, $\cos 2 \phi_R$,
the two other observables are quartic in these variables. A potential
ambiguity in the determination of $(\cos 2 \phi_L, \cos 2 \phi_R)$ will be, 
however, very unlikely, especially if $m_{\tilde{\nu}}$ is fixed
independently and the c.m. energy varied.
We do not dwell here on further aspects of step (I) which
can be found in \cite{CDDKZ}.\\

We now go to step (II) of eq.(1) and describe briefly how to determine
$M_2, \mu$ and  $\tan\beta$. Starting from eq.(\ref{mixing}) and
$m_{\chi_1^+}$ in eq.(\ref{chmass}), CDDKZ give closed expressions for 
$M_2, \mu, \tan\beta$ in terms of the quantities 
$p= \cot (\phi_R - \phi_L), q= \cot (\phi_R + \phi_L)$. They considered all
possible cases and concluded to the existence of at most a two-fold ambiguity
in the determination of the Lagrangian parameters, traceable to a sign ambiguity
in $\sin 2 \phi_{L, R}$, (see \cite{CDDKZ} for details). Here we only
sketch an equivalent discussion  which shows that, when it occurs, this
two-fold ambiguity is 
{\sl always associated with opposite $\mu$ sign solutions}.
This can be most easily seen as follows: from $\cos 2 \phi_L, \cos 2 \phi_R$
in eq.(\ref{mixing}) one easily determines $M_2$ uniquely (remember that
$M_2$ is positive in our convention) and $\mu$ with a global sign
ambiguity, as functions of $m_{\chi^+_1}, \tan \beta, \cos 2 \phi_L$
and $\cos 2 \phi_R$. Plugging those functions in the $m_{\chi^+_1}$
part of eq.(\ref{chmass}) one gets, thanks to some cancellations,
a simple quadratic equation in $\tan \beta$. The two solutions encompass
automatically the sign of $\sin 2 \phi_L \sin 2 \phi_R$. Furthermore, each
of them is consistent only with (at most) one $\mu$ sign reproducing
the correct $m_{\chi^+_1}$, since eq.(\ref{chmass}) is not invariant
under $ \mu \to - \mu$ for a given $M_2, \tan \beta$.
As a numerical illustration, taking the input of fig.2,
$\sigma_{tot} = 0.37$pb, ${\cal P}^2/{\cal Q} = -0.24,
{\cal P}^2/{\cal Y} = -3.66$,   
CDDKZ find the following two-fold solution
\begin{eqnarray}
 [\tan\beta; M_2,\mu] =
        \left\{\begin{array}{l}
              (A) \;\;[1.06; \;  83{\rm GeV}, \; -59{\rm GeV}] \\
                  { }\\ {}
              (B) \;\;[3.33; \;  248{\rm GeV}, \; 123{\rm GeV}]
              \end{array}\right.
\label{numillust}
\end{eqnarray}
We see that the two-fold ambiguity comes with a sign change for $\mu$
in accord with the general pattern just described. To eliminate this discrete
ambiguity one would clearly need an independent information about any of the 
three parameters. Finally, the reconstruction obviously depends on the quality
of the experimental accuracy with which the needed observables can be
determined. This requires among other things:
\begin{itemize}
\item[-] running at different c.m. energies: at threshold for a good
determination of $m_{\chi^+_1}$, away from threshold to increase the sensitivity
to chargino polarization;
\item[-] a good reconstruction of the final state fermion systems for a
good determination of the neutralino mass;
\item[-] identification of the chargino electric charge, necessary for the
extraction of ${\cal P}$;
\item[-] an independent knowledge of the sneutrino mass,  to avoid a
three parameter fit to the observables;
\end{itemize}
 
\subsection{KM}

In this section we describe another strategy for extracting the
Lagrangian parameters \cite{KM}. It consists in assuming that only ino 
masses are known.
Among other things, this strategy will be complementary
to the one described in the previous section in the sense that it provides 
(within the CDDKZ strategy) an
algorithm for the determination of $M_1$, the only parameter which was not
reconstructed in ref.[1]. KM concerns mainly step (II) of eq.(1). The
emphasis is put on the extent to which the reconstruction can be made through
a controllable analytical procedure including all possible ambiguities, if three
ino masses and $\tan \beta$ were known\footnote{The fact that $\tan \beta$
needs to be an input is actually a marginal point here, as one can assume that
this parameter has been determined from elsewhere, like for instance in 
\cite{CDDKZ} or from the study of yet another sector of the MSSM.}.
This is particularly relevant for the neutralino sector where the analytical 
reconstruction is far from trivial.

The next aim in \cite{KM} is to provide a numerical code which uses as much
of the analytical solutions as possible and allows a direct reconstruction
of $M_1, M_2$ and $\mu$ from the physical ino masses. We do not address here
the more realistic issues when only mass differences are measured 
\cite{detection}, however
it is clear that the study provides a useful building block even in this
case, and practically allows to avoid parameter scanning numerical procedures 
as well as model-dependent assumptions.  KM distinguish two cases:
\begin{itemize}
\item[$S_1$:] The two charginos and one neutralino masses are input;
\item[$S_2$:] One chargino and two neutralino masses are input;
\end{itemize}

Although $S_1$ is phenomenologically less compelling than $S_2$ as far as
the generic pattern of low lying states is concerned, it turns out
that it leads to a full analytical reconstruction. In contrast, $S_2$
needs partly a purely numerical algorithm which is, however, minimized through
the use of the $S_1$ solutions. The bottom line is that the resulting algorithms
are very fast, the first being fully analytical and the second needing
seldom more than a few iterations to reach numerical convergence   
(see \cite{KM} for more details).\\
Let us now describe briefly the solutions for $S_1$.\\
\underline{\sl{Chargino sector:}}\\
Starting from eq.(\ref{chmass}) one can determine analytically
$\mu^2$ and $M_2$ in terms of $M_{\chi^+_1}, M_{\chi^+_2}, \tan \beta$
and $m_W$. Without further information in the chargino sector,
$\mu$ and $M_2$ will be determined, but up to a $|\mu| \leftrightarrow M_2$ 
ambiguity (that is, one cannot determine uniquely at this level the Higgsino 
and Gaugino content of the charginos). On the other hand
the global sign ambiguity in $\mu$, due to the fact that only $\mu^2$ is known, is actually lifted by the relation
\begin{equation}
\label{sgnmu}
M_2 \;\mu = m^2_W \sin 2\beta \pm 
M_{\chi^+_1} M_{\chi^+_2}\;
\end{equation}
since $M_2$ is positive by definition. Nonetheless there remains
a two-fold ambiguity coming from the relative $\pm$ sign in eq.(\ref{sgnmu}).
On the other hand, some constraints will come from the requirement of
real-valuedness of $M_2$ and $\mu$. All these aspects are analytically
delineated in \cite{KM} in terms of domains of $\tan \beta$ and the sum and
difference of the input chargino masses.\\
\underline{\sl {Neutralino sector:}}\\
Let us now assume that $M_2, \mu, \tan \beta$ and one neutralino mass have been
determined, and address the question of reconstructing $M_1$ and thus the three 
remaining neutralino masses. It should be clear that the answer to this question
is not straightforward independently of whether it can be phrased analytically
or not. Indeed, with all parameters but $M_1$ fixed in eq.(\ref{neutmass}), and
the knowledge of the mass of just one neutralino state (say the lightest), 
it could well be that multiple branch solutions exist which would be lifted 
only through extra information about the couplings in this sector. It turns out, however, not to be the case (at least when phases are ignored):
there is basically a unique solution, apart from
the fact that one should allow for negative and positive values for the
input neutralino mass since ${\cal M_N}$ can have negative eigenvalues.
(This sign liberty is actually the only ambiguity which can be eventually
fixed through the study of the couplings and will not be discussed further here.)\\
The trick is to write down the four independent combinations of the entries
of ${\cal M_N}$ which are invariant under similarity transformations, and
thus relate them simply to the four eigenvalues of ${\cal M_N}$.
One can then express the correlations between the eigenvalues and the basic 
parameters in the following form:
     
\begin{equation}
\begin{array}{lll}
 M_1&=& 
   -\frac{P_{i j}^2 + 
       P_{i j} (\mu^2 + m_Z^2 + M_2 S_{i j} - S_{i j}^2) 
              + \mu m_Z^2 M_2 s_w^2 \sin 2 \beta}
      {P_{i j} (S_{i j} -M_2) + \mu ( c_w^2 m_Z^2 \sin 2 \beta -\mu M_2 )}
\end{array}\label{solM1}
\end{equation}

\begin{equation}
\begin{array}{lll}
M_2 &=& 
\frac{ S_{i j} P_{i j}^2 + P_{i j} m_Z^2 \mu \sin 2 \beta
 -( P_{i j}^2 + (\mu^2 + m_w^2) P_{i j} + S_{i j} m_w^2 \mu \sin 2 \beta) M_1 }{
P_{i j}^2 + P_{i j} ( \mu^2 + s_w^2 m_Z^2) + 
 \mu S_{i j} ( s_w^2 m_Z^2 \sin 2 \beta
- \mu M_1) }  
\end{array}\label{condition2}
\end{equation}

where
\begin{eqnarray} 
S_{i j} \equiv \tilde{M}_{N_i} + \tilde{M}_{N_j} \nonumber \\
P_{i j} \equiv \tilde{M}_{N_i}  \tilde{M}_{N_j} \nonumber \\
\nonumber
\end{eqnarray}
and $i\neq j$ index any neutralino mass parameter.
(The tilde denotes the fact that the mass parameters can be negative valued)
The nice thing about the above equations is that if any of the neutralino
masses is taken as input (say  $\tilde{M}_{N_2}$), then the other three
are determined analytically through a simple cubic equation. 
A unique value for $M_1$ is then determined from eq.(\ref{solM1}) after 
plugging any of these solutions. Eqs.(\ref{solM1}, \ref{condition2}) express
in a specially convenient way the various correlations among the four
eigenvalues and the basic parameters. It is also noteworthy that the
input set $(M_2, \mu, \tan \beta)$ {\sl plus} one neutralino mass is optimal
for a fully analytical determination. In particular
this is precisely the input set required in ref.[1]. 
We illustrate here the complementarity of the two approaches by taking the 
two sets of numbers (A) and (B) in eq.(\ref{numillust}) and a lightest
neutralino $M_{N_1}= 30 GeV$,
to reconstruct $M_1$ and the remaining neutralino masses 
from eqs.(\ref{solM1}, \ref{condition2}):

\begin{equation}
\begin{array}{l}
\; \; \; \; \;\;\;\;\; \; \; \; \;\;\;\;
[\;\; M_1,\;\;\;\;\;\;\tilde{M}_{N_2},\;\;\;\;\;\; 
\tilde{M}_{N_3},\;\;\;\;\;\;\;\;\;\tilde{M}_{N_4}] 
{}\\{}\\
(A) 
        \left\{\begin{array}{l}
  (+);\; [ 30{\rm GeV}, \;  59{\rm GeV}, \; -107{\rm GeV}, \; 122{\rm GeV}] \\
                  { }\\ {}
  (-);\; [ -52{\rm GeV}, \;  58{\rm GeV}, \; -119{\rm GeV}, \; 120{\rm GeV}]
          \end{array}\right. \\
                  { }\\ {}
(B)
        \left\{\begin{array}{l}
  (+);\; [ 46{\rm GeV}, \;  110{\rm GeV}, \; -130{\rm GeV}, \; 284{\rm GeV}] \\
                  { }\\ {}
  (-);\; [ -25{\rm GeV}, \; 101{\rm GeV}, \; -132{\rm GeV}, \; 284{\rm GeV}]
                        \end{array}\right.
\label{numillust1}
\end{array}
\end{equation}  

Here $(\pm)$ refer to the two possible signs of the $30 GeV$ lightest
eigenmass input. The effect of this sign tends to be more important for $M_1$
than for the neutralino masses. Of course a minus sign should be accompanied
with the appropriate sign change in the Feynman rules involving
neutralinos (see \cite{haber+erratum}). A further study of the left and
right form factors in the chargino decay system could then lift partially
the four-fold ambiguity in eq.(\ref{numillust1}). Lifting the remaining two-fold
ambiguity will still necessitate further measurements from the ino sector.\\

Back to the $S_1$ strategy, we give 
in fig.3 an illustration of the sensitivity to a chargino mass,
fixing the other two masses of chargino and neutralino. 
The behavior of the reconstructed $(M_1, M_2, \mu)$ turns out to be fairly 
simple, up to the two-fold ambiguity induced by $\mu$ in the chargino sector.
A simple behavior shows as well for the remaining three neutralino masses
(see \cite{KM}), 
when the input neutralino mass is varied. This behavior which is fully 
controlled analytically can be used to discuss the generic gross features of 
the spectrum even when one deviates from the present input strategy. 
For instance the sensitivity to $\tan \beta$ turns out to be rather mild, and 
the effect of the sign change in the input neutralino mass tends to be 
negligible apart from well localized regions (see ref.[2] for further 
illustrations, including a reconstruction of the parameters at the
GUT scale).\\
In fig.4 we illustrate the $S_2$ strategy. The input set in this case
requires a partial numerical algorithm since the output becomes controlled
by high degree polynomials. However using eqs.(\ref{solM1}, \ref{condition2})
in conjunction with the chargino sector relations allows an optimized
iterative algorithm.  Fig.4 shows a rather intricate behavior of
$(M_1, M_2, \mu)$ when one chargino and two neutralino masses are taken
as input, a reflection of the above mentioned high degree polynomials,
which nevertheless boils down (at least in our numerical trials) to at
most a four-fold ambiguity. The regions of many-fold ambiguities 
or no ambiguity at all are separated by domains where the output parameters
become complex valued (the shaded areas). Furthermore the singular behavior in 
some small regions is generically traced  back to zeroing some parameters 
(see ref.[2] for more details). 

\begin{figure}
\center
%\rule{2cm}{0.2mm}\hfill \rule{2cm}{0.2mm}
%\vskip 4cm
%\rule{2cm}{0.2mm}\hfill \rule{2cm}{0.2mm}
%\epsfig{figure=cont.eps,height=1.5in}
\epsfig{figure=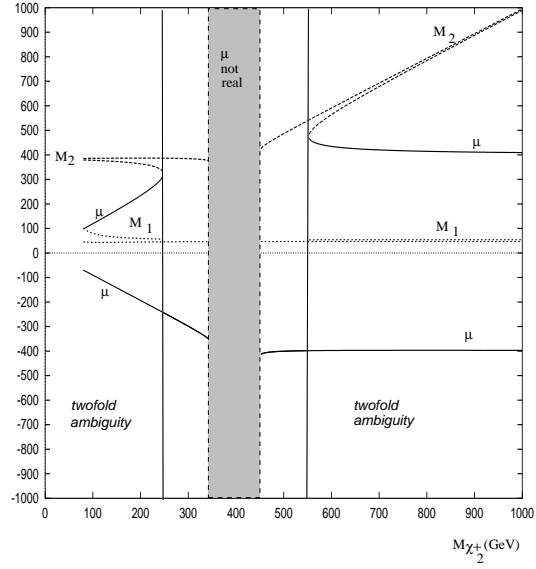, height= 3in}
\caption{$\mu, M_1$ and $M_2$ (with the ``higgsino-like'' convention $|\mu| \leq M_2$) as functions of $M_{\chi^+_2}$ for fixed $M_{\chi^+_1}$ ( = 400 GeV);  
$M_{N_2}$ ( = 50 GeV), and $\tan\beta$ ( = 2). (taken from ref.[2]) }
\label{fig:km1}
\end{figure}

\begin{figure}
\center
%\rule{2cm}{0.2mm}\hfill \rule{2cm}{0.2mm}
%\vskip 4cm
%\rule{2cm}{0.2mm}\hfill \rule{2cm}{0.2mm}
%\epsfig{figure=cont.eps,height=1.5in}
\epsfig{figure=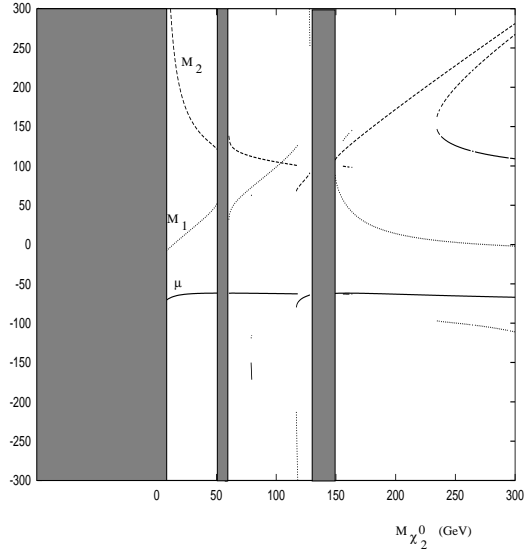, height= 3in}
\caption{\label{fig5} $\mu$, $M_2$ and $M_1$ as function of  
$M_{\chi_2}^0$ 
for fixed $M_{N_3}$ ( $= -100 $GeV), $M_{\chi^+_1}$ 
(= 80 GeV)  
and $\tan\beta( = 2$). (taken from ref.[2])}
\label{fig:km2}
\end{figure}

\section{Final comments}
In this talk we presented two possible theoretical strategies for the extraction
of the ino sector parameters from physical observables. The first relied on the
study of the production and decay of the lightest chargino in $e^+ e^-$ 
collisions, the second on the knowledge of some ino masses.
We also illustrated a full reconstruction of the ino sector when the two
complementary approaches are brought together. Generally speaking, 
these approaches provide with efficient tools for the study
of the ino sector. In the same time, 
they suggest the need in some cases for further experimental information
due to the occurrence of possible discrete ambiguities in the reconstruction.\\
Furthermore, although we only considered real-valued parameters, 
some of the material
presented here goes through unaltered if phases are allowed.
This is the case in CDDKZ for the chargino sector, 
even though extra information will still be needed
to determine those phases. The inclusion of phases is less obvious in KM, 
especially in the neutralino sector, and deserves a separate study by itself.
One should, however, keep in mind that the above strategies can give
indirect information about the need for phases, whenever the experimental
data place the parameters in the forbidden regions delineated in KM.
In any case, a by-product of the analytical study would have been
the construction of fast and flexible algorithms which can be used
in various ways when reconstructing the ino parameters. 

Finally, it should be stressed that the strategies we presented here are just
at the theoretical level. Obviously a more realistic examination of the 
experimental extraction of observables and related errors is still needed
to assess their degree of efficiency. In addition, these strategies should
eventually be placed in a wider context involving the other sectors of the MSSM,
taking into account plausible discovery scenarios of the susy partners. 
The inter-correlations between these sectors, endemic to supersymmetry, will 
then hopefully allow a unique determination of all the parameters of 
the model.  

\section*{Acknowledgements}
I would like to thank Jean-Lo\"{\i}c Kneur and Peter Zerwas for valuable 
discussions on their contributed papers. Thanks go also to 
Seongyoul Choi and Abdelhak Djouadi for providing me with useful material
for the talk.

%\section*{References}

\end{document}